%% file: main.tex
\documentclass[submission,copyright,creativecommons]{eptcs}
\providecommand{\event}{QPL 2019}
\usepackage{hyperref}
\hypersetup{
    colorlinks,
    linkcolor={red!50!black},
    citecolor={blue!50!black},
    urlcolor={blue!80!black}
}

\input{preamble.tex}

\usepackage{rwd-drafting}

\def\titlerunning{PyZX: Large Scale Automated Diagrammatic Reasoning}
\title{\titlerunning}

\author{Aleks Kissinger
\institute{University of Oxford}
\email{aleks.kissinger@cs.ox.ac.uk} \and
John van de Wetering
\institute{Radboud University}
\email{john@vdwetering.name}
}
\def\authorrunning{A. Kissinger \& J.~van de Wetering}

\begin{document}
\maketitle

\begin{abstract}
    The ZX-calculus is a graphical language for reasoning about ZX-diagrams, a type of tensor networks that can represent arbitrary linear maps between qubits. Using the ZX-calculus, we can intuitively reason about quantum theory, and optimise and validate quantum circuits. In this paper we introduce PyZX, an open source library for automated reasoning with large ZX-diagrams. We give a brief introduction to the ZX-calculus, then show how PyZX implements methods for circuit optimisation, equality validation, and visualisation and how it can be used in tandem with other software. We end with a set of challenges that when solved would enhance the utility of automated diagrammatic reasoning.
\end{abstract}

\section{Introduction}

The ZX-calculus was introduced in 2008 by Coecke and Duncan~\cite{CD1,CD2}. Since then, much progress has been made on understanding this language: rule-sets have been found that make the language complete~\cite{SimonCompleteness,HarnyAmarCompleteness,euler-zx}; variants of the language have been developed that allow easy reasoning about different domains, e.g.\ ZW for quantum entanglement~\cite{CK,hadzihasanovic2017algebra} as well as Fermionic quantum computing~\cite{hadzihasanovic2018diagrammatic} and ZH for Toffoli-Hadamard circuits~\cite{backens2018zhcalculus}; and a diverse set of applications have been found including measurement-based quantum computing~\cite{DP2,kissinger2017MBQC}, surface code lattice surgery~\cite{horsman2017surgery} and circuit optimisation~\cite{duncan2019graph,optimisation-paper}. However, in order to bring these applications into the practical domain, we need tools that allow us to reason with ZX-diagrams on a large scale.

PyZX (pronounced like `physics' without the H) is a Python-based library designed for reasoning with large quantum circuits and ZX-diagrams. PyZX is Free and Open Source Software, licensed under GPLv3. The project is hosted on GitHub and available at:
\begin{center}
  \href{https://github.com/Quantomatic/pyzx}{\texttt{https://github.com/Quantomatic/pyzx}}
\end{center}
It allows users to efficiently rewrite ZX-diagrams using built-in simplification strategies. These strategies are powerful enough to reduce Clifford diagrams to a pseudo-normal form~\cite{duncan2019graph}, perform highly effective T-count optimisation~\cite{optimisation-paper}, and to verify correctness of optimised circuits.
Besides the powerful set of rewrite strategies, PyZX offers many ways to present and visualise circuits and ZX-diagrams. It can input and output circuits in the QASM, QC, and Quipper formats, and it can convert ZX-diagrams into TikZ diagrams or the Quantomatic qGraph format.

\begin{figure}[!bht]
  \centering
  \scalebox{1.2}{\input{schematic.tikz}}
  \caption{An overview of the functionality of PyZX.}\label{fig:schematic}
\end{figure}
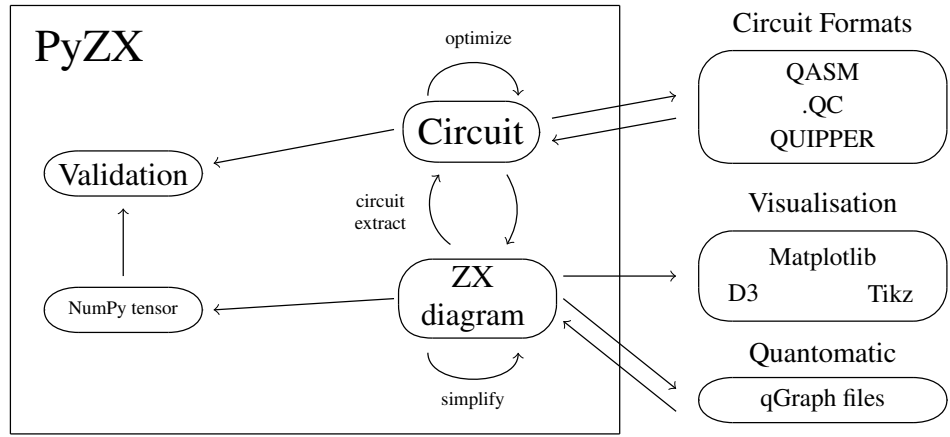

PyZX is best within Jupyter Notebooks~\cite{jupyter}, a popular way to run Python interactively with rich input and output features. In this setting ZX-diagrams can be visualised on the fly. PyZX can also be called as a command-line tool for circuit-to-circuit optimisation, e.g. for optimising many circuits or combining with other procedures. See Figure~\ref{fig:schematic} for an overview.

As much of the theory behind PyZX has already been detailed elsewhere~\cite{duncan2019graph,optimisation-paper}, we will focus in this paper on the PyZX tool itself. We describe the capabilities of PyZX and several ways in which PyZX can be used, and we give some of the implementation details. We end the paper by offering several challenges to the community, each of which when completed would expand the power of automated diagrammatic reasoning, or would help in understanding its limitations.

We start by giving a brief overview of the ZX-calculus in Section~\ref{sec:zxcalculus}. Then, in Section~\ref{sec:core-functionality} we describe the core functionality of PyZX and give some details about how it is implemented. In Section~\ref{sec:tikzit-quanto} we detail how PyZX interacts with TikZ and Quantomatic. We discuss the performance of PyZX in Section~\ref{sec:performance}, and finally, in Section~\ref{sec:challenges} we present a series of challenges for ourselves and other potential contributors/collaborators.


\section{The ZX-calculus}\label{sec:zxcalculus}

This section provides a brief overview of the \zxcalculus. For an in-depth reference see Ref.~\cite{CKbook}.

The \zxcalculus is a diagrammatic language similar to the familiar
quantum circuit notation.  A \emph{\zxdiagram} (or simply
\emph{diagram}) consists of \emph{wires} and \emph{spiders}.  Wires
entering the diagram from the left are \emph{inputs}; wires exiting to
the right are \emph{outputs}.  Given two diagrams we can compose them
by joining the outputs of the first to the inputs of the second, or
form their tensor product by simply stacking the two diagrams.

Spiders are linear operations which can have any number of input or output
wires.  There are two varieties: $Z$ spiders depicted as green dots and $X$ spiders depicted as red dots:\footnote{If you are reading this document in monochrome or otherwise have difficulty distinguishing green and red, $Z$ spiders will appear lightly-shaded and $X$ darkly-shaded.}
\[
\hfill \input{Zsp-a.tikz} \ := \ \ketbra{0...0}{0...0} +
e^{i \alpha} \ketbra{1...1}{1...1} \hfill
\qquad
\hfill \input{Xsp-a.tikz} \ := \ \ketbra{+...+}{+...+} +
e^{i \alpha} \ketbra{-...-}{-...-} \hfill
\]
The diagram as a whole corresponds to a linear map built from the
spiders (and SWAPs represented by wire crossings) by the usual composition and tensor product
of linear maps.  As a special case, diagrams with no inputs represent
(unnormalised) state preparations.

It will be convenient to introduce a symbol -- a yellow square -- for
the Hadamard gate. This is defined by the equation:
\begin{equation}\label{eq:Hdef}
\hfill
\input{had-alt.tikz}
\hfill
\end{equation}

We will not view this Hadamard gate as an additional generator.
Instead, we consider a Hadamard box between two spiders as a different type of edge.
We illustrate this by writing a Hadamard between two spiders as a blue dashed edge:
\begin{equation}\label{eq:def-had-edge}
\input{blue-edge-def.tikz} 
\end{equation}
Both the blue edge notation and the Hadamard box can always be
translated back into spiders when necessary. We will refer to the blue
edge as a \emph{Hadamard edge}.

Two diagrams are considered \emph{equal} when one can be deformed to
the other by moving the vertices around in the plane, bending,
unbending, crossing, and uncrossing wires, so long as the connectivity
and the order of the inputs and outputs is maintained.  

\begin{figure}[b!]
\centering
\begin{tabular}{|c|}
\hline \\[-0.2cm]
\quad \scalebox{0.93}{\input{ZX-rules.tikz}} \quad\ \\[2.3cm]
\hline 
\end{tabular}
\caption{\label{fig:zx-rules} A convenient presentation for the ZX-calculus. These rules hold
  for all $\alpha, \beta \in [0, 2 \pi)$, and due to $(\bm{h})$ and
  $(\bm{i2})$ all rules also hold with the colours
  interchanged.}
\end{figure}
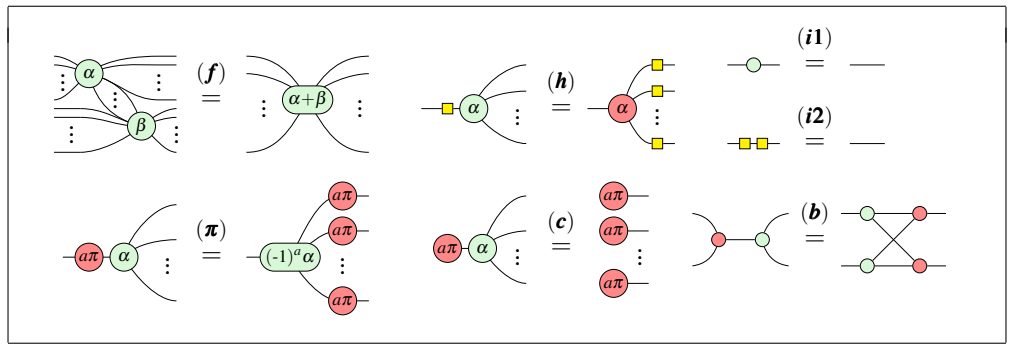

Additionally, there is a set of equations that we call the \emph{rules} of the \zxcalculus that allow us to equate additional diagrams; these are shown in
Figure~\ref{fig:zx-rules}. Equal diagrams correspond to equal linear maps. In this paper we will ignore global scalar factors, and hence we will consider linear maps $A$ and $B$ to be ``equal'' if they satisfy $A = zB$ for some non-zero $z\in\C$. The rules of the calculus can be made exact with respect to scalar factors~\cite{Backens:2015aa}, but we do not consider this possibility in this paper.

Quantum circuits can be translated into ZX-diagrams in a straightforward manner. 
We will take as our starting point circuits constructed
from the following universal set of gates:
\[
\text{CNOT} \ :=\
\left(\begin{matrix}
  1 & 0 & 0 & 0 \\
  0 & 1 & 0 & 0 \\
  0 & 0 & 0 & 1 \\
  0 & 0 & 1 & 0 \\
\end{matrix}\right)
\qquad\qquad
Z_\alpha \ :=\
\left(\begin{matrix}
  1 & 0 \\
  0 & e^{i \alpha}
\end{matrix}\right)
\qquad\qquad
H \ :=\ \frac{1}{\sqrt{2}}
\left(\begin{matrix}
  1 & 1 \\
  1 & -1
\end{matrix}\right)
\]
This gate set allows a convenient representation in terms of spiders: 
\begin{align}\label{eq:zx-gates}
\text{CNOT} & = \input{cnot.tikz} &
Z_\alpha & = \input{Z-a.tikz} &
H & = \input{h-alone.tikz}
\end{align}
Note that the green spider is the first (i.e. control) qubit of the CNOT and the red spider is the second (i.e. target) qubit.
Other common gates can easily be expressed in terms of these gates. In particular, $S := Z_{\frac\pi2}$, $T := Z_{\frac\pi4}$ and:
\begin{align}\label{eq:zx-derived-gates}
X_\alpha & = \input{X-a-expanded.tikz} &
\text{CZ} & = \input{cz-small.tikz}
\end{align}

When we want to write a quantum circuit as a ZX-diagram, we simply convert each of the gates of the circuit into their equivalent ZX-diagram, and compose these per the structure of the circuit. For circuits using gates not described above, such as the Toffoli gate, we first convert the gate into a Clifford+T representation, and then we write it as a ZX-diagram.

\section{Core functionality}\label{sec:core-functionality}

In this section we detail the main features of PyZX and how they are implemented. Our intention is to give a brief tour of the capabilities of PyZX, rather than to give a detailed tutorial. Full documentation is available at \href{https://pyzx.readthedocs.io}{\tt pyzx.readthedocs.io}.

PyZX is written in pure Python with the only dependencies being NumPy~\cite{numpy} and Matplotlib~\cite{matplotlib}. A relatively easy way to get these dependencies (and Jupyter) is to use Anaconda~\cite{anaconda}.

\subsection{ZX-diagrams and circuits}\label{sec:circuit-graph}

There are two main data-structures present in PyZX: {\circuit}s and {\graph}s. A \circuit is basically a wrapper around a list of gates, while a \graph represents a ZX-diagram. 

The \circuit class is the entry-point for importing and exporting circuits to and from PyZX. It also provides methods to do gate-level operations, such as converting a Toffoli-circuit into a Clifford+T circuit or taking the adjoint of all its gates. There is also a variety of circuit optimisation schemes that act directly on the \circuit class which can be found in the \texttt{optimize} sub-module. 

A \circuit consists of a list of {\gate}s, which in turn are small classes containing some information about the gate and how to convert it into various representations, such as ZX-diagrams or the QASM~\cite{qasmpaper} format.

The \graph class is more interesting. The graphs in PyZX are simple, undirected graphs with typed vertices and edges. Vertices come in three types: boundaries, Z-spiders and X-spiders. Each vertex can be labelled by a phase that is stored as a fraction representing a rational multiple of $\pi$. For instance, a label of $\frac12$ corresponds to a phase of $\frac\pi2$. 

The edges come in two types, which correspond respectively to a regular connection between spiders, and a Hadamard-edge as shown in Eq.~\eqref{eq:def-had-edge}. As ZX-diagrams allow parallel edges between spiders and self-loops, we need a way to deal with these in PyZX graphs. Adding an edge where there is already one present (via the \texttt{add\_edge} method) will simply replace it. Often, it is more convenient to use the rules of the ZX-calculus to resolve parallel edges and self-loops whenever a new edge is added. How this should be done depends on the types of edges and vertices involved:
\ctikzfig{edge-cases}
In order to use this built-in functionality to resolve the edges, one can use the \graph method \texttt{add\_edge\_table}, which takes in a list of edges and edge-types, and resolves what the ZX-diagram would look like when all the double edges and self-loops have been resolved.


The \graph class abstracts away some of the details of the underlying representation of the graph. As a result, a different implementation of a graph can simply subclass the \texttt{BaseGraph} class in order to work with all the other functionality of PyZX. 

The default, pure Python, implementation of a graph in PyZX, \texttt{GraphS}, stores its connectivity as a dictionary of dictionaries, where the first level has vertices as keys (identified by an integer), with the values being another dictionary containing all the neighbours of this vertex. The values of these dictionaries in turn specify by which type of vertex the vertices are connected. Phases are stored in another dictionary. We have experimented with other graph back-ends, but we found that this simple implementation is fast enough to handle and simplify diagrams with hundreds of thousands of vertices in a reasonable time-frame. See Section~\ref{sec:performance} for more details. 

\begin{figure}[!htb]
    \centering
    \includegraphics[width=0.9\textwidth]{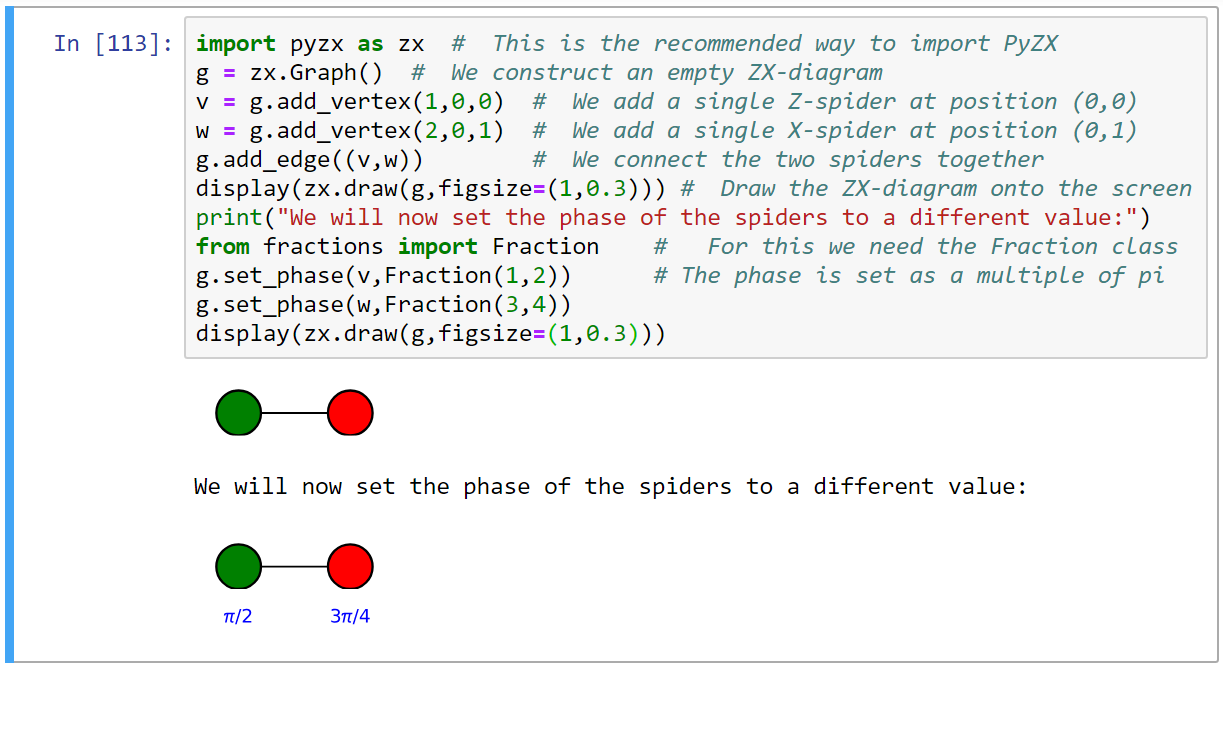}
    \vspace{-0.8cm}
    \caption{Simple example of construction a ZX-diagram from scratch and visualising it.}\label{fig:simple-example}
\end{figure}

Figure~\ref{fig:simple-example} gives a small example of creating a ZX-diagram in PyZX and visualising it. This example takes place inside a Jupyter Notebook~\cite{jupyter}. A Jupyter Notebook is an interactive Python shell that runs in a web-browser and allows one to easily rerun parts of your code, and to add interactive HTML and Javascript widgets. We use Jupyter as a GUI to PyZX, using its integration with Matplotlib to visualise ZX-diagrams.

Usually you will not want to manually create a ZX-diagram. There are various ways to import ZX-diagrams into PyZX. Directly importing ZX-diagrams is possible using the Quantomatic format (see Section~\ref{sec:tikzit-quanto}), but PyZX can also read files describing quantum circuits in a variety of languages. It currently supports QASM~\cite{qasmpaper}, the Quipper ASCII format~\cite{quipper} and the QC/TFC format used by the Reversible Circuit Benchmarks page~\cite{maslovreversible}. A circuit can be imported by calling for instance
 \texttt{circ = zx.Circuit.load("circuit.extension")}. The file-format is automatically inferred. As the circuit might contain gates that have no straightforward translation to ZX-diagrams, we do not automatically convert the imported circuit into a ZX-diagram. The method 
 \texttt{to\_graph()} transforms a \circuit into a \graph by converting the circuit into a Clifford+phases representation that is easily translated to a ZX-diagram.

\subsection{Simplification and Verification}
The simplification strategies for ZX-diagrams in PyZX are built in a hierarchical manner. 

At the bottom level there are \emph{rules} that consist of a \emph{matcher} and a \emph{rewriter}. A matcher loops over all the vertices or edges of a graph (which one depends on the rule) and tries to find as many non-overlapping sections of the graph that are suitable for the rule to be applied. Once it finishes, it returns a list of matches (what such a match object looks like differs per rule). The rewriter takes this list of matches and figures out which changes to the graph need to be made. Since all the matches are non-overlapping, this can be done for each match separately. If these changes involve adding vertices or changing phases, then this is done immediately, but in order to reduce overhead, other changes (adding edges or removing edges and vertices) are collected until the rewriter has processed all matches, and then implemented simultaneously. 

For a simple example of a rule, let us consider \emph{identity removal}:
\begin{equation}\label{eq:id-simp}
\input{id-simp.tikz}
\end{equation}
The matcher loops over vertices and tries to find those vertices that have a phase of zero, and exactly two neighbours. When it has found such a vertex, it removes its neighbours from the list of candidates (in order to prevent overlapping applications), and it makes it into a match that contains the vertex, its neighbours, and the type of the edge that should be made between these neighbours in the rewritten graph (in the case above, since both edges to the middle vertex were Hadamard edges, the resulting edge is a regular edge, cf.~$(\bm{i2})$). The rewriter takes this list of matches, builds a list of vertices to be removed, and edges to be added, and applies this all at once. As the neighbours of the removed vertex could already have been connected in the original graph, this could lead to a self-loop that is handled in the way described in Section~\ref{sec:circuit-graph}.

The next stage in the hierarchy of simplification strategies are the \emph{basic simplifiers}. These are built on a single rule. A simplifier keeps applying the matcher and rewriter of a single rule, until the matcher finds no more matches. Because a matcher only finds non-overlapping rules, and thereby might miss possible applications for the rule the first time around, and because the rewritten diagram might produce new sections of the graph that are suitable for application of the rule, the simplifier might need to do the process of matching and rewriting many times before no new matches are found. In order for the simplifier to not get stuck in an infinite loop, it is important that the rule actually simplifies the diagram in some manner, i.e.\ that some kind of metric on the graph is reduced. In the case of the basic simplifiers used in PyZX, this metric is usually that the amount of vertices in the graph is reduced, but it could also be something more complex such as trading one type of vertex for another one.

Let us consider the identity removal simplifier (named \texttt{id\_simp} in PyZX). Its simplifier removes all the identity spiders in the diagram as in \eqref{eq:id-simp}. The rule never generates new arity 2 zero phase spiders, hence the only way in which the simplifier has to run multiple times is when there were multiple identities in a row, i.e.\ when the rule applications had overlap. Since every application of the rule removes a vertex, this simplifier indeed terminates.

At the top level of the hierarchy are the \emph{compound simplifiers}. These simply combine other simplifiers into a more complicated simplification strategy, by applying other simplifiers in some particular order, potentially looping over this combination until none of the simplifiers finds any more reductions.

For instance, PyZX also implements a basic simplifier that fuses spiders together (cf.~$(\bm{f})$), called \texttt{spider\_simp}. We could combine \texttt{id\_simp} and \texttt{spider\_simp} together in a compound simplifier as follows:
{
\begin{python}
def fuse_simp(g):
    i = 0
    while True:
        i1 = id_simp(g)     # simplifiers return the amount of iterations
        i2 = spider_simp(g) 
        if i1 == 0 and i2 == 0: break  # No more matches were found
        i += 1
    return i
\end{python}
}
This simplifier simply keeps removing identities and fusing spiders until this is no longer possible. Note that removing an identity might make more possibilities for fusing spiders, and fusing spiders might make a zero phase arity two spider. Hence, we indeed need the loop in this function to make full use of these simplifiers.

\begin{figure}[!htb]
    \centering
    \includegraphics[width=0.9\textwidth]{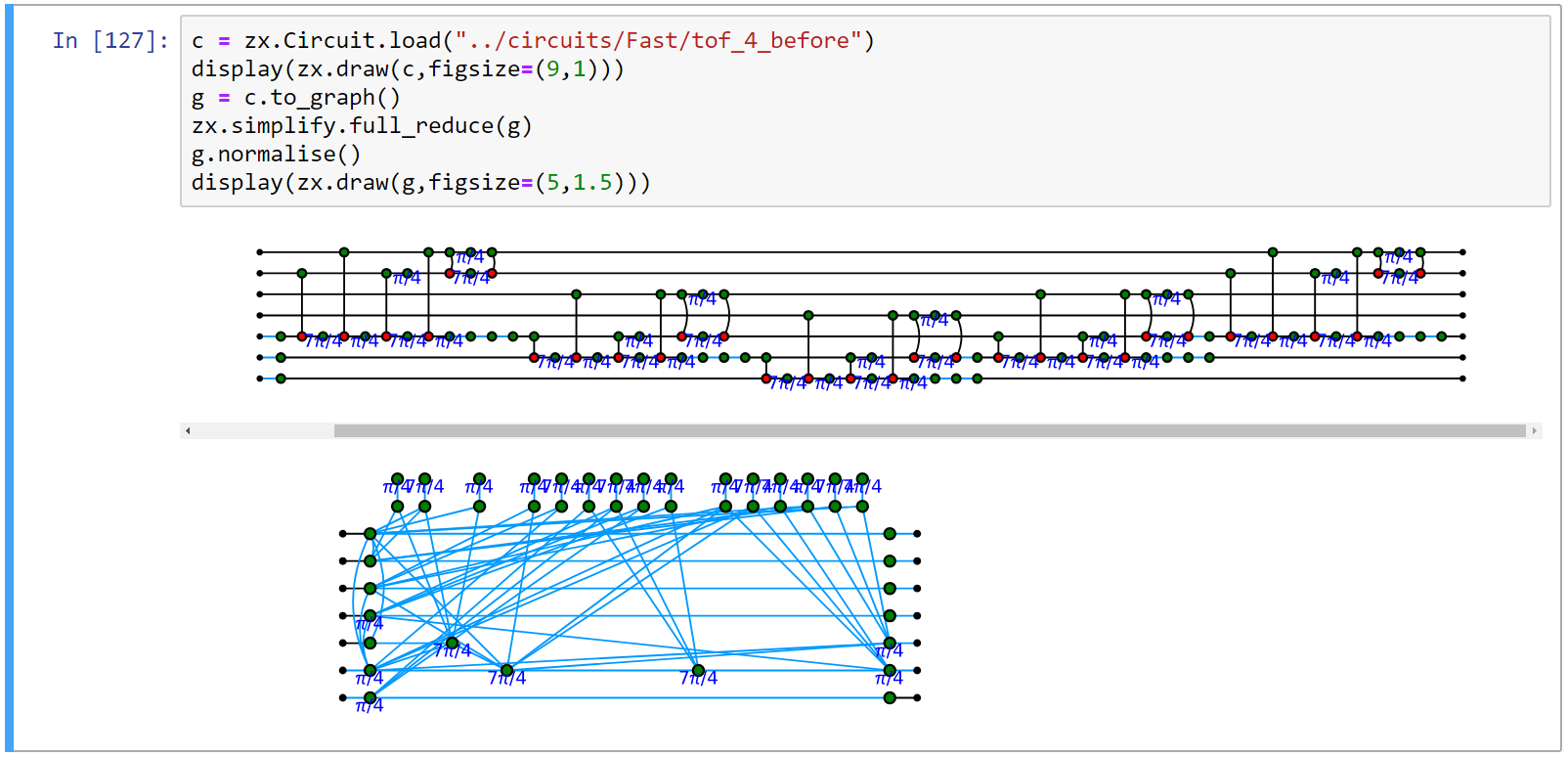}
    \caption{Example of simplifying a circuit with PyZX. The circuit \texttt{tof\_4\_before} is from Ref.~\cite{nam2018automated}. The function \texttt{g.normalise()} does not change the ZX-diagram, but changes the coordinates of the vertices in order to make it more suitable for drawing. Note that the blue wires of the bottom diagram correspond to the Hadamard edges of \eqref{eq:def-had-edge}.}\label{fig:simp}
\end{figure}

The rewrite strategies of PyZX are included in the \texttt{simplify} sub-package. It includes the previously described simplification strategies, but also more complex rewrite strategies based on the \emph{local complementation} and \emph{pivoting} procedures of Ref.~\cite{duncan2019graph}. These rewrite strategies are combined into two top level compound rewrite strategies: \texttt{zx.simplify.clifford\_simp} implements the rewrite strategy of Ref.~\cite{duncan2019graph} that is able to reduce Clifford circuits to GS-LC (graph-state with local Cliffords) pseudo-normal form, while \texttt{zx.simplify.full\_reduce} builds on that by implementing the strategy of Ref.~\cite{optimisation-paper} that is able to do significant T-count optimisation. For a demonstration of the usage of the rewrite system see Figure~\ref{fig:simp}.

It is of course essential to know whether the rewrite rules are implemented correctly and actually preserve the semantics of the diagram. PyZX offers a straightforward way to do this by allowing ZX-diagrams to be converted into their underlying linear maps using NumPy~\cite{numpy}. To test equality of diagrams we can then simply test whether all the elements in the two tensors are equal up to a global non-zero scalar:
\begin{python}
    t1 = g1.to_tensor()
    t2 = g2.to_tensor()
    zx.compare_tensors(t1,t2)
\end{python}
This method uses an exponential amount of memory in terms of the qubits, or in the case of general ZX-diagrams: some function of the amount of inputs and the maximal arity of the spiders involved. In practice it can validate circuits up to 10 qubits on a 32-bit version of Python.

Interestingly, we can also use the rewriting engine itself to validate equality of diagrams or circuits. Given circuits (or just any ZX-diagrams) \texttt{c1} and \texttt{c2} we construct the circuit \texttt{c} as:
\begin{python}
c = c1.adjoint()
c.add_circuit(c2)
\end{python}
If \texttt{c1} and \texttt{c2} indeed implement the same unitary, then \texttt{c} should implement the identity. To test the equality of \texttt{c} to the identity, we try to simplify it as much as possible. If we can successfully reduce it to the identity circuit, then we know (at least with very high likelihood) that \texttt{c} is indeed the identity. The function \texttt{c1.verify\_equality(c2)}implements this idea using the strategy of Ref.~\cite{optimisation-paper} (i.e.\ \texttt{zx.simplify.full\_reduce}). This method is fast (for circuits with dozens of qubits and thousands of gates, it finishes in seconds), and if equality is verified, it could in principle even give a certificate of this equality in terms of a list of rewrite rules. But if this method fails to reduce the circuit to the identity, then the answer is inconclusive, so that the utility of this method is determined by the variety of circuits it can verify.

We have run this validation method on the original and optimised circuits of Ref.~\cite{nam2018automated} and managed to validate equality of all pairs of circuits except for one, where it turned out that the optimised circuit contained an error.
Based on the ability of \texttt{full\_reduce} to validate correctness of all these circuits, we speculate that whenever two ZX-diagrams can be transformed into one another by a combination of the rules of Figure~\ref{fig:zx-rules}, that \texttt{full\_reduce} should be able to verify equality of these circuits.

\subsection{Circuit-to-circuit optimisation}

As can be seen in Figure~\ref{fig:simp}, when a circuit-like ZX-diagram is fed into the simplification procedure, the simplified diagram might no longer resemble the structure of a circuit. If one is interested in using these simplifications for circuit optimisation, the resulting diagram should then be converted back into a circuit somehow. This is known as the \emph{circuit extraction problem}~\cite{duncan2019graph}. To convert a ZX-diagram into a circuit in PyZX one can use the function \texttt{zx.extract.streaming\_extract}. This method uses the proven results regarding circuit extraction from Ref.~\cite{duncan2019graph}, but builds on that with a series of additional heuristics to extract a circuit from a ZX-diagram. There is currently no results known about when these heuristics suffice to extract a circuit, but when the diagram is produced from a circuit by \texttt{zx.simplify.full\_reduce} then the extraction always seems to succeed\footnote{We have tested the extraction on thousands of randomly generated circuits of varying sizes, next to dozens of standard benchmark circuits. In every case, our extraction heuristic successfully terminates.}. Alternatively, PyZX also implements the \emph{phase teleportation} routine from Ref.~\cite{optimisation-paper} that allows one to optimise circuits `in place', side-stepping the issue of circuit extraction. This method is implemented in \texttt{zx.simplify.teleport\_reduce}.

Next to these optimisation methods that use the ZX-calculus, PyZX also implements some straight circuit-to-circuit optimisation methods. In particular, it contains an implementation of the TODD phase-polynomial optimiser of Ref.~\cite{heyfron2018efficient}.

Instead of doing these operations manually inside a Python shell, PyZX also offers a command-line interface, e.g.\ by calling \texttt{python -m pyzx opt input\_circuit}.

\section{TikZiT and Quantomatic integration}\label{sec:tikzit-quanto}

Besides the core functionality based on the rewrite engine, PyZX offers a couple of convenience features for interacting with other software. Any ZX-diagram in PyZX can be exported as a TikZ figure for easy importing into a Latex document. This is how many of the bigger diagrams in Ref.~\cite{duncan2019graph} were produced. The TikZ output is compatible with the TikZ figure editing software TikZiT~\cite{tikzit} and in fact, PyZX allows direct export into TikZiT:
\begin{python}
    zx.tikz.tikzit_location = 'path/to/tikzit.executable'
    zx.tikz.tikzit(g)  #  This opens Tikzit with the ZX-diagram g loaded
\end{python}

Alternatively, a circuit can be converted to a TikZ ZX-diagram by a command-line call: \texttt{python -m pyzx tikz input\_circuit}.

PyZX also integrates with the graphical theorem prover Quantomatic~\cite{kissinger2015quantomatic} in various ways. The file-format that Quantomatic uses to store diagrams, \emph{qGraph}, is supported by PyZX, allowing non-circuit ZX-diagrams to be stored and read. Furthermore, when Quantomatic is installed, it can be used as a GUI for editing ZX-diagrams:
\begin{python}
zx.quantomatic.quantomatic_location = '/path/to/quantomatic.jar'
g2 = zx.quantomatic.edit_graph(g)  #  Open Quantomatic with diagram g loaded
# The above function blocks until Quantomatic is closed
# Now g2 contains whatever changes were made to g in Quantomatic
\end{python}

The rewrite strategies of PyZX are also available from inside Quantomatic, by importing PyZX as a Python module. For more details, we refer to the documentation~\cite{pyzxdoc}.

\section{Performance}\label{sec:performance}

\begin{figure}[!htb]
\centering
  \includegraphics[width=0.9\textwidth]{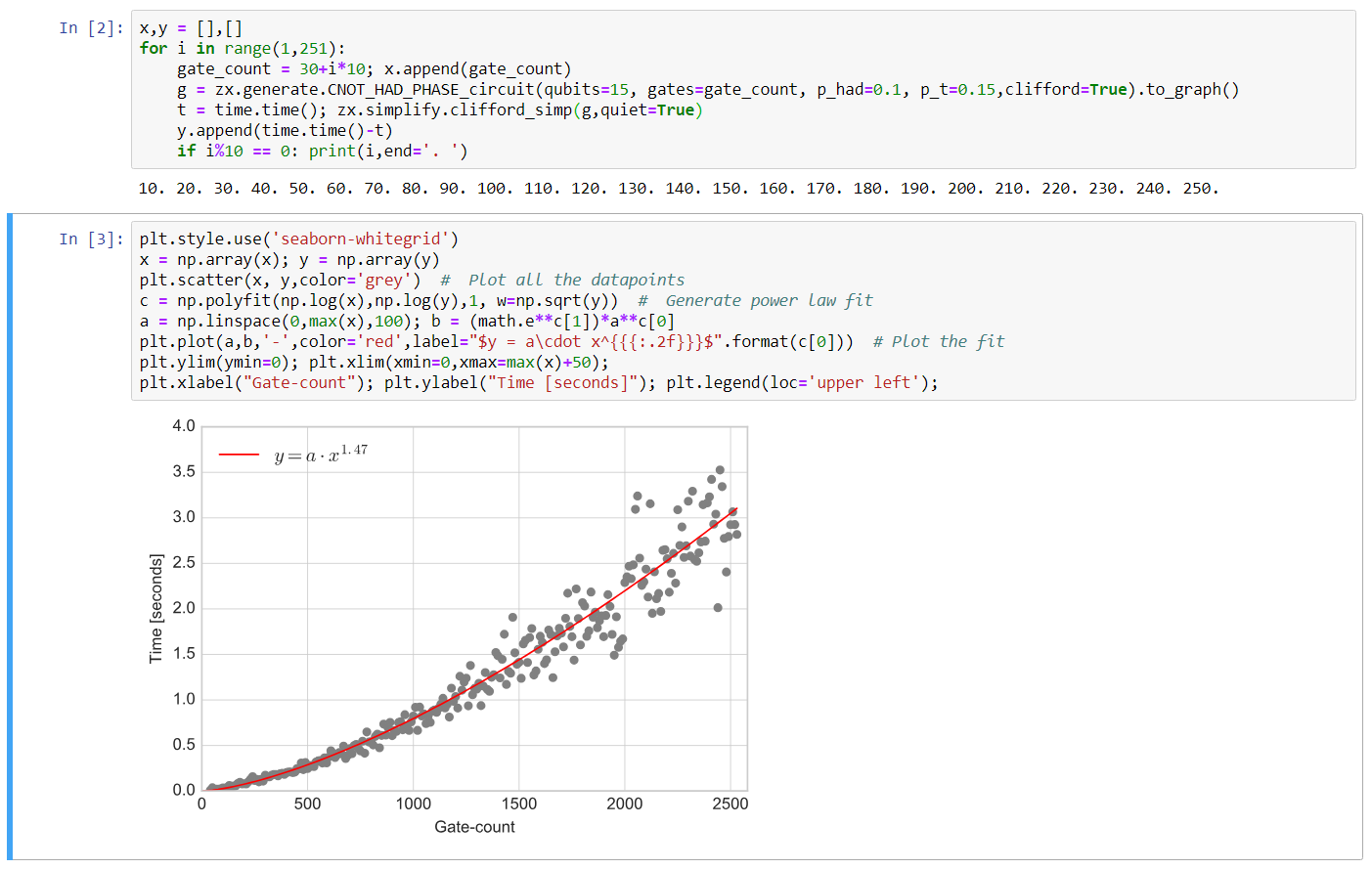}
  \caption{Benchmark of the Clifford simplification procedure that reduces circuits to normal form, as implemented in a Jupyter Notebook. The red-line is the least squares best-fit for a power-law. This was run on a consumer laptop.}\label{fig:benchmark}
\end{figure}

PyZX is able to handle and simplify ZX-diagrams containing tens of thousands of vertices in seconds; See Figure~\ref{fig:benchmark}. As can be seen, reducing a 15 qubit circuit with 2500 gates takes less then 5 seconds. Interestingly, the time it takes to bring a Clifford circuit to normal form seems to follow a power-law, although the constant of the power-law depends on the amount of qubits in the circuit. 

Let us compare this to a, on the surface, similar piece of software: Quantomatic. Like PyZX, Quantomatic supports automated reasoning with ZX-diagrams~\cite{kissinger2015quantomatic}. There are however two important differences. 

The first is that Quantomatic is designed to be agnostic concerning the particular graphical language that is being used. As a result, particular behaviour that is `hard-coded' into PyZX, such as the removal of parallel edges between vertices, needs to be implemented as a separate rule in Quantomatic, slowing down the way in which graphs can be simplified. 

The second difference is that the rewrite engine of Quantomatic is based on \emph{bang-boxes}~\cite{kissinger2012pattern}. Whenever a rule allows a variable amount of edges or vertices to be involved, it must be implemented in Quantomatic using bang-boxes. While a powerful tool, finding matches involving bang-boxes can be slow. As every rule in PyZX is a piece of hand-written Python code, it does not have to rely on the strict logic of bang-boxes. While this necessitates more validation to make sure the implementation is correct, the result is that PyZX can be much faster. 

In Ref.~\cite{FaganDuncan} Quantomatic was used to optimise Clifford circuits. Optimising a 5 qubit circuit with less than 100 gates took upwards of half an hour. Contrast this to Figure~\ref{fig:benchmark} that shows PyZX doing the similar task of bringing Clifford circuits with tens of qubits and thousands of gates to normal-form in seconds.

Of course, simplifying a Clifford circuit to normal-form is not a standard task for most purposes, but even for the more common use of optimising the T-count of a circuit, PyZX is fast. For instance, the implementation of the phase teleportation routine of Ref.~\cite{optimisation-paper} can optimise most of the circuits used in Ref.~\cite{nam2018automated} in seconds. For example, the 9 qubit circuit \texttt{nth\_prime6.tfc} from the Reversible Circuit Benchmark page~\cite{maslovreversible} contains 1241 gates, 567 of which are T gates. PyZX optimises this to a circuit with 279 T gates in less than a second.

\section{Challenges}\label{sec:challenges}


While PyZX is already a capable, general-purpose circuit optimiser~\cite{optimisation-paper}, there are still many ways in which PyZX can be improved. We list some of these as a series of challenges. 
\medskip

\noindent \textbf{Challenge 1: Use ancillae in optimising ZX-diagrams}. While progress has been made on optimising circuits using the ZX-calculus~\cite{duncan2019graph,optimisation-paper}, these results only concern ancilla-free optimisation. It is currently still unknown how the ZX-calculus might be used to create optimisations that take advantage of the presence of ancillae.
\medskip

\noindent \textbf{Challenge 2: Find the limitations of circuit extraction.} Understanding of the circuit extraction problem is still limited~\cite{duncan2019graph}. Even though we have a heuristic that always seems to work in the diagrams we produce by simplification, better understanding of the extraction problem might lead to ways to extract smaller circuits, which would be beneficial to optimisation. We believe the general problem of circuit extraction to be hard (in the complexity-theoretic manner), so finding specific circumstances in which we can successfully extract a circuit is of both theoretical and practical importance.
\medskip

\noindent \textbf{Challenge 3: Combine circuit routing with the ZX-calculus.} Circuits on physical quantum computers (and even logical level descriptions of circuits such as the surface code) do not allow two-qubit interactions to take place between any pair of qubits. If a two-qubit gate needs to take place between qubits that are not adjacent, then some form of \emph{routing} needs to happen. When extracting a circuit from a ZX-diagram, there is a natural place for circuit routing to happen, as noted in Ref.~\cite{kissinger2019cnot}. It would be interesting to see whether this leads to improvements over other methods. 
\medskip

\noindent \textbf{Challenge 4: Use the ZH-calculus for circuit optimisation.} The ZX-calculus is very well-suited for describing circuits with Clifford and phase gates. It can however only indirectly reason about Toffoli and CCZ gates. A more natural candidate for describing those gates is the \emph{ZH-calculus}~\cite{backens2018zhcalculus}. It stands to reason then that the optimisation of these circuits can be improved by using the ZH-calculus.
\medskip

\noindent \textbf{Challenge 5: Use graphical calculi for circuit simulation.} An intriguing possibility for the usage of ZX-diagrams, is in \emph{strong simulation} of quantum circuits, i.e.\ determining specific output probabilities given a specific input state. Such a problem can be described by a tensor network~\cite{markov2008simulating}, or as a scalar ZX-diagram. Using the simplification engine of PyZX, simplifications to the diagram can potentially make the calculation of the scalar more efficient then in other approaches (a similar approach to simplifying tensor networks has already shown that this can make the calculation significantly more efficient~\cite{chen2018classical}). %
\medskip

\noindent \textbf{Challenge 6: Use the ZX-calculus for lattice surgery compilation.} The ZX-calculus is a natural choice for describing lattice surgery on a surface code~\cite{horsman2017surgery}. As a result, one can imagine there being some kind of automated way to convert a quantum circuit into a sequence of lattice surgery operations that can be performed on a surface code.
\medskip

\noindent \textbf{Acknowledgements}: The authors are supported in part by AFOSR grant FA2386-18-1-4028. The authors wish to thank Will Zeng for supporting the development of PyZX with a Unitary Fund, and Arianne Meijer for developing code to do circuit routing in PyZX.

\bibliographystyle{eptcs}
\bibliography{main}

\end{document}

%% file: preamble.tex
\usepackage[utf8]{inputenc}
\usepackage{scrextend}
\usepackage[english]{babel}
\usepackage{amsthm}
\usepackage{amsfonts}
\usepackage{amsmath}
\usepackage{stmaryrd}
\usepackage{graphicx}
\usepackage{keycommand}
\usepackage{hyperref}
\usepackage[all]{hypcap}

\usepackage{listings}
\DeclareFixedFont{\ttb}{T1}{txtt}{bx}{n}{10} 
\DeclareFixedFont{\ttm}{T1}{txtt}{m}{n}{10}  
\usepackage{xcolor}
\definecolor{deepblue}{rgb}{0,0,0.5}
\definecolor{deepred}{rgb}{0.6,0,0}
\definecolor{deepgreen}{rgb}{0,0.5,0}
\newcommand\pythonstyle{\lstset{
language=Python,
basicstyle=\ttm,
otherkeywords={self},             
keywordstyle=\ttb\color{deepblue},
emph={MyClass,__init__},          
emphstyle=\ttb\color{deepred},    
stringstyle=\color{deepgreen},
frame=tb,                         
showstringspaces=false,
commentstyle=\color{gray} %
}}

\lstnewenvironment{python}[1][]
{
\pythonstyle
\lstset{#1}
}
{}
\newcommand\pythoninline[1]{{\pythonstyle\lstinline!#1!}}

\theoremstyle{definition}

\newtheorem{example*}[theorem]{Example*}
\newtheorem{examples*}[theorem]{Examples*}

\newtheorem{remark*}[theorem]{Remark*}

\newtheorem*{theorem*}{Theorem}
\newtheorem*{corollary*}{Corollary}
\newtheorem*{lemma*}{Lemma}
\newtheorem*{proposition*}{Proposition}

\usepackage{tikzit}
\input{zx.tikzdefs}
\input{zx.tikzstyles}

\makeatletter
\newcommand\etc{etc\@ifnextchar.{}{.\@}\xspace}

\makeatother

\usepackage{bm}

\newcommand{\bra}[1]{\ensuremath{\left\langle #1 \right|}}
\newcommand{\ket}[1]{\ensuremath{\left|  #1 \right\rangle}}

\newcommand{\ketbra}[2]{\ensuremath{\ket{#1}\!\bra{#2}}}
\newcommand{\C}{\mathbb{C}}

\newcommand{\circuit}{\texttt{Circuit}\xspace}
\newcommand{\graph}{\texttt{Graph}\xspace}
\newcommand{\gate}{\texttt{Gate}\xspace}

%% file: zx.tikzstyles
\tikzstyle{box}=[shape=rectangle, text height=1.5ex, text depth=0.25ex, yshift=0.5mm, fill=white, draw=black, minimum height=5mm, yshift=-0.5mm, minimum width=5mm, font={\small}]

\tikzstyle{Z dot}=[inner sep=0mm, minimum size=2mm, shape=circle, draw=black, fill={rgb,255: red,216; green,248; blue,216}]
\tikzstyle{Z phase dot}=[minimum size=5mm, font={\footnotesize\boldmath}, shape=rectangle, rounded corners=2mm, inner sep=0.2mm, outer sep=-2mm, scale=0.8, tikzit shape=circle, draw=black, fill={rgb,255: red,216; green,248; blue,216}, tikzit draw=blue]
\tikzstyle{X dot}=[Z dot, shape=circle, draw=black, fill={rgb,255: red,255; green,136; blue,136}]
\tikzstyle{X phase dot}=[Z phase dot, tikzit shape=circle, tikzit draw=blue, fill={rgb,255: red,255; green,136; blue,136}, font={\footnotesize\color{black}\boldmath}]

\tikzstyle{hadamard}=[fill=yellow, draw=black, shape=rectangle, inner sep=0.6mm, minimum height=1.5mm, minimum width=1.5mm]
\tikzstyle{vertex}=[inner sep=0mm, minimum size=1mm, shape=circle, draw=black, fill=black]
\tikzstyle{vertex set}=[inner sep=0mm, minimum size=1mm, shape=circle, draw=black, fill=white, font={\footnotesize\boldmath}]

\tikzstyle{hadamard edge}=[-, color=blue, dashed, dash pattern=on 3pt off 1.5pt, thick]
\tikzstyle{brace edge}=[-, tikzit draw=blue, decorate, decoration={brace,amplitude=1mm,raise=-1mm}]
\tikzstyle{diredge}=[->]

%% file: figures/schematic.tikz
\begin{tikzpicture}
	\begin{pgfonlayer}{nodelayer}
		\node [style=none] (0) at (-10.75, 8.75) {};
		\node [style=none] (1) at (-10.75, -0.75) {};
		\node [style=none] (2) at (2.75, -0.75) {};
		\node [style=none] (3) at (2.75, 8.75) {};
		\node [style=none] (4) at (-9, 7.75) {\Large PyZX};
		\node [style=none] (5) at (-0.55, 5.95) {\large Circuit};
		\node [style=none] (6) at (-0.5, 2.25) {\parbox{1.7cm}{\small\centering ZX\\diagram}};
		\node [style=none] (7) at (-1, 3.45) {};
		\node [style=none] (8) at (-1.25, 5) {};
		\node [style=none] (9) at (-2.55, 4.175) {\parbox{1.4cm}{\tiny\centering circuit\\extract}};
		\node [style=none] (10) at (0.25, 5) {};
		\node [style=none] (11) at (0.25, 3.45) {};
		\node [style=none] (16) at (7.25, 8.375) {\footnotesize Circuit Formats};
		\node [style=none] (17) at (7.25, 7.25) {\scriptsize QASM};
		\node [style=none] (18) at (7.25, 6.5) {\scriptsize.QC};
		\node [style=none] (19) at (7.25, 5.75) {\scriptsize QUIPPER};
		\node [style=none] (20) at (7.25, 4.375) {\footnotesize Visualisation};
		\node [style=none] (25) at (7.25, 3.125) {\scriptsize Matplotlib};
		\node [style=none] (26) at (5.5, 2.375) {\scriptsize D3};
		\node [style=none] (27) at (8.75, 2.375) {\scriptsize Tikz};
		\node [style=none] (28) at (-1.5, 6.75) {};
		\node [style=none] (29) at (0.5, 6.75) {};
		\node [style=none] (30) at (-0.375, 8) {\tiny optimize};
		\node [style=none] (31) at (-1.5, 1.05) {};
		\node [style=none] (32) at (0.5, 1.05) {};
		\node [style=none] (33) at (-0.5, 0) {\tiny simplify};
		\node [style=none] (34) at (4, 2.75) {};
		\node [style=none] (35) at (1.5, 2.75) {};
		\node [style=none] (36) at (1.25, 6.25) {};
		\node [style=none] (37) at (4, 6.75) {};
		\node [style=none] (38) at (4, 6.25) {};
		\node [style=none] (39) at (1.25, 5.75) {};
		\node [style=none] (44) at (-8.25, 5) {\small Validation};
		\node [style=none] (46) at (-2.25, 6) {};
		\node [style=none] (47) at (-6.25, 5.25) {};
		\node [style=none] (48) at (-2.25, 2.25) {};
		\node [style=none] (49) at (-6.25, 2) {};
		\node [style=none] (50) at (7.25, 1) {\footnotesize Quantomatic};
		\node [style=none] (55) at (7.25, 0) {\scriptsize qGraph files};
		\node [style=none] (56) at (1.5, 2.25) {};
		\node [style=none] (57) at (4, 0.25) {};
		\node [style=none] (58) at (4, -0.25) {};
		\node [style=none] (59) at (1.5, 1.75) {};
		\node [style=none] (60) at (-8.25, 2) {\tiny NumPy tensor};
		\node [style=none] (61) at (-8.25, 2.75) {};
		\node [style=none] (62) at (-8.25, 4.25) {};
		\node [style=none] (67) at (4.5, 7) {};
		\node [style=none] (68) at (5.25, 7.75) {};
		\node [style=none] (69) at (4.5, 6) {};
		\node [style=none] (70) at (5.25, 5.25) {};
		\node [style=none] (71) at (9.25, 7.75) {};
		\node [style=none] (72) at (10, 7) {};
		\node [style=none] (73) at (10, 6) {};
		\node [style=none] (74) at (9.25, 5.25) {};
		\node [style=none] (78) at (4.5, 3) {};
		\node [style=none] (79) at (5.25, 3.75) {};
		\node [style=none] (80) at (4.5, 2.5) {};
		\node [style=none] (81) at (5.25, 1.75) {};
		\node [style=none] (82) at (9.25, 3.75) {};
		\node [style=none] (83) at (10, 3) {};
		\node [style=none] (84) at (10, 2.5) {};
		\node [style=none] (85) at (9.25, 1.75) {};
		\node [style=none] (89) at (4.5, 0) {};
		\node [style=none] (90) at (5.25, 0.5) {};
		\node [style=none] (91) at (4.5, 0) {};
		\node [style=none] (92) at (5.25, -0.5) {};
		\node [style=none] (93) at (9.25, 0.5) {};
		\node [style=none] (94) at (10, 0) {};
		\node [style=none] (95) at (10, 0) {};
		\node [style=none] (96) at (9.25, -0.5) {};
		\node [style=none] (98) at (-10, 2) {};
		\node [style=none] (99) at (-9.25, 2.5) {};
		\node [style=none] (100) at (-10, 2) {};
		\node [style=none] (101) at (-9.25, 1.5) {};
		\node [style=none] (102) at (-7.25, 2.5) {};
		\node [style=none] (103) at (-6.5, 2) {};
		\node [style=none] (104) at (-6.5, 2) {};
		\node [style=none] (105) at (-7.25, 1.5) {};
		\node [style=none] (118) at (-2.15, 2.375) {};
		\node [style=none] (119) at (-1.4, 3.125) {};
		\node [style=none] (120) at (-2.15, 2.125) {};
		\node [style=none] (121) at (-1.4, 1.375) {};
		\node [style=none] (122) at (0.6, 3.125) {};
		\node [style=none] (123) at (1.35, 2.375) {};
		\node [style=none] (124) at (1.35, 2.125) {};
		\node [style=none] (125) at (0.6, 1.375) {};
		\node [style=none] (127) at (-2.05, 6.025) {};
		\node [style=none] (128) at (-1.25, 6.625) {};
		\node [style=none] (129) at (-2.05, 6) {};
		\node [style=none] (130) at (-1.25, 5.275) {};
		\node [style=none] (131) at (0.225, 6.625) {};
		\node [style=none] (132) at (0.975, 5.925) {};
		\node [style=none] (133) at (0.975, 5.95) {};
		\node [style=none] (134) at (0.225, 5.275) {};
		\node [style=none] (137) at (-10, 5.025) {};
		\node [style=none] (138) at (-9.25, 5.525) {};
		\node [style=none] (139) at (-10, 5.025) {};
		\node [style=none] (140) at (-9.25, 4.525) {};
		\node [style=none] (141) at (-7.25, 5.525) {};
		\node [style=none] (142) at (-6.5, 5.025) {};
		\node [style=none] (143) at (-6.5, 5.025) {};
		\node [style=none] (144) at (-7.25, 4.525) {};
	\end{pgfonlayer}
	\begin{pgfonlayer}{edgelayer}
		\draw (0.center) to (3.center);
		\draw (3.center) to (2.center);
		\draw (2.center) to (1.center);
		\draw (0.center) to (1.center);
		\draw [style=diredge, bend left=45] (7.center) to (8.center);
		\draw [style=diredge, bend left] (10.center) to (11.center);
		\draw [style=diredge, bend left=90, looseness=1.25] (28.center) to (29.center);
		\draw [style=diredge, bend right=90] (31.center) to (32.center);
		\draw [style=diredge] (35.center) to (34.center);
		\draw [style=diredge] (36.center) to (37.center);
		\draw [style=diredge] (38.center) to (39.center);
		\draw [style=diredge] (46.center) to (47.center);
		\draw [style=diredge] (56.center) to (57.center);
		\draw [style=diredge] (58.center) to (59.center);
		\draw [style=diredge] (48.center) to (49.center);
		\draw [style=diredge] (61.center) to (62.center);
		\draw [bend left=45] (67.center) to (68.center);
		\draw [bend right=45, looseness=1.25] (69.center) to (70.center);
		\draw (69.center) to (67.center);
		\draw (70.center) to (74.center);
		\draw [bend right=45] (74.center) to (73.center);
		\draw (73.center) to (72.center);
		\draw [bend right=45] (72.center) to (71.center);
		\draw (71.center) to (68.center);
		\draw [bend left=45] (78.center) to (79.center);
		\draw [bend right=45, looseness=1.25] (80.center) to (81.center);
		\draw (80.center) to (78.center);
		\draw (81.center) to (85.center);
		\draw [bend right=45] (85.center) to (84.center);
		\draw (84.center) to (83.center);
		\draw [bend right=45] (83.center) to (82.center);
		\draw (82.center) to (79.center);
		\draw [in=-180, out=90] (89.center) to (90.center);
		\draw [in=-180, out=-90] (91.center) to (92.center);
		\draw (91.center) to (89.center);
		\draw (92.center) to (96.center);
		\draw [in=-90, out=0] (96.center) to (95.center);
		\draw (95.center) to (94.center);
		\draw [in=0, out=90] (94.center) to (93.center);
		\draw (93.center) to (90.center);
		\draw [in=-180, out=90] (98.center) to (99.center);
		\draw [in=-180, out=-90] (100.center) to (101.center);
		\draw (100.center) to (98.center);
		\draw (101.center) to (105.center);
		\draw [in=-90, out=0] (105.center) to (104.center);
		\draw (104.center) to (103.center);
		\draw [in=0, out=90] (103.center) to (102.center);
		\draw (102.center) to (99.center);
		\draw [bend left=45] (118.center) to (119.center);
		\draw [bend right=45, looseness=1.25] (120.center) to (121.center);
		\draw (120.center) to (118.center);
		\draw (121.center) to (125.center);
		\draw [bend right=45] (125.center) to (124.center);
		\draw (124.center) to (123.center);
		\draw [bend right=45] (123.center) to (122.center);
		\draw (122.center) to (119.center);
		\draw (129.center) to (127.center);
		\draw (130.center) to (134.center);
		\draw (133.center) to (132.center);
		\draw (131.center) to (128.center);
		\draw [in=90, out=180, looseness=1.25] (128.center) to (127.center);
		\draw [in=180, out=-90, looseness=1.25] (129.center) to (130.center);
		\draw [in=-90, out=0] (134.center) to (133.center);
		\draw [in=90, out=0] (131.center) to (132.center);
		\draw [in=-180, out=90] (137.center) to (138.center);
		\draw [in=-180, out=-90] (139.center) to (140.center);
		\draw (139.center) to (137.center);
		\draw (140.center) to (144.center);
		\draw [in=-90, out=0] (144.center) to (143.center);
		\draw (143.center) to (142.center);
		\draw [in=0, out=90] (142.center) to (141.center);
		\draw (141.center) to (138.center);
	\end{pgfonlayer}
\end{tikzpicture}

%% file: figures/Zsp-a.tikz
\begin{tikzpicture}
	\begin{pgfonlayer}{nodelayer}
		\node [style=Z phase dot] (0) at (0, 0) {$\alpha$};
		\node [style=none] (1) at (1.25, 1) {};
		\node [style=none] (2) at (-1.25, 1) {};
		\node [style=none] (3) at (-1.25, -1) {};
		\node [style=none] (4) at (1.25, -1) {};
		\node [style=none] (5) at (1.25, 0.5) {};
		\node [style=none] (6) at (-1.25, 0.5) {};
		\node [style=none, rotate=90] (7) at (-1, -0.25) {...};
		\node [style=none, rotate=90] (8) at (1, -0.25) {...};
	\end{pgfonlayer}
	\begin{pgfonlayer}{edgelayer}
		\draw [in=-141, out=0, looseness=0.75] (3.center) to (0);
		\draw [in=180, out=-39, looseness=0.75] (0) to (4.center);
		\draw [in=180, out=22, looseness=0.75] (0) to (5.center);
		\draw [in=180, out=39, looseness=0.75] (0) to (1.center);
		\draw [in=0, out=158, looseness=0.75] (0) to (6.center);
		\draw [in=141, out=0, looseness=0.75] (2.center) to (0);
	\end{pgfonlayer}
\end{tikzpicture}

%% file: figures/Xsp-a.tikz
\begin{tikzpicture}
	\begin{pgfonlayer}{nodelayer}
		\node [style=X phase dot] (0) at (0, 0) {$\alpha$};
		\node [style=none] (1) at (1.25, 1) {};
		\node [style=none] (2) at (-1.25, 1) {};
		\node [style=none] (3) at (-1.25, -1) {};
		\node [style=none] (4) at (1.25, -1) {};
		\node [style=none] (5) at (1.25, 0.5) {};
		\node [style=none] (6) at (-1.25, 0.5) {};
		\node [style=none, rotate=90] (7) at (-1, -0.25) {...};
		\node [style=none, rotate=90] (8) at (1, -0.25) {...};
	\end{pgfonlayer}
	\begin{pgfonlayer}{edgelayer}
		\draw [in=-141, out=0, looseness=0.75] (3.center) to (0);
		\draw [in=180, out=-39, looseness=0.75] (0) to (4.center);
		\draw [in=180, out=22, looseness=0.75] (0) to (5.center);
		\draw [in=180, out=39, looseness=0.75] (0) to (1.center);
		\draw [in=0, out=158, looseness=0.75] (0) to (6.center);
		\draw [in=141, out=0, looseness=0.75] (2.center) to (0);
	\end{pgfonlayer}
\end{tikzpicture}

%% file: figures/had-alt.tikz
\begin{tikzpicture}
	\begin{pgfonlayer}{nodelayer}
		\node [style=none] (0) at (-2.75, 0) {};
		\node [style=none] (1) at (-0.75, 0) {};
		\node [style=hadamard] (2) at (-1.75, 0) {};
		\node [style=none] (3) at (0, 0) {$=$};
		\node [style=none] (4) at (0.75, 0) {};
		\node [style=Z phase dot] (5) at (1.5, 0) {$\frac\pi2$};
		\node [style=X phase dot] (6) at (2.5, 0) {$\frac\pi2$};
		\node [style=Z phase dot] (7) at (3.5, 0) {$\frac\pi2$};
		\node [style=none] (8) at (4.25, 0) {};
	\end{pgfonlayer}
	\begin{pgfonlayer}{edgelayer}
		\draw (0.center) to (2);
		\draw (2) to (1.center);
		\draw (4.center) to (8.center);
	\end{pgfonlayer}
\end{tikzpicture}

%% file: figures/blue-edge-def.tikz
\begin{tikzpicture}
	\begin{pgfonlayer}{nodelayer}
		\node [style=none] (0) at (-4.5, 0) {};
		\node [style=none] (1) at (-3, 1) {};
		\node [style=none] (2) at (-1.75, -1.5) {};
		\node [style=none] (3) at (-1.25, 0.5) {};
		\node [style=Z dot] (4) at (-3.75, 0.5) {};
		\node [style=Z dot] (5) at (-1.75, -0.5) {};
		\node [style=none] (6) at (-4, 1.5) {};
		\node [style=none] (7) at (-3, -1) {};
		\node [style=none] (8) at (0, 0) {$:=$};
		\node [style=hadamard] (17) at (2.75, -0.25) {};
		\node [style=none] (18) at (2.5, 0.75) {};
		\node [style=none] (19) at (3.75, -1.75) {};
		\node [style=Z dot] (20) at (3.75, -0.75) {};
		\node [style=Z dot] (21) at (1.75, 0.25) {};
		\node [style=none] (22) at (1.5, 1.25) {};
		\node [style=none] (23) at (2.5, -1.25) {};
		\node [style=none] (24) at (4.25, 0.25) {};
		\node [style=none] (25) at (1, -0.25) {};
	\end{pgfonlayer}
	\begin{pgfonlayer}{edgelayer}
		\draw (3.center) to (5);
		\draw (5) to (2.center);
		\draw (0.center) to (4);
		\draw (4) to (1.center);
		\draw (7.center) to (5);
		\draw (6.center) to (4);
		\draw [style=hadamard edge] (4) to (5);
		\draw (24.center) to (20);
		\draw (20) to (19.center);
		\draw (25.center) to (21);
		\draw (21) to (18.center);
		\draw (23.center) to (20);
		\draw (22.center) to (21);
		\draw (21) to (17);
		\draw (17) to (20);
	\end{pgfonlayer}
\end{tikzpicture}

%% file: figures/ZX-rules.tikz
\begin{tikzpicture}
	\begin{pgfonlayer}{nodelayer}
		\node [style=Z phase dot] (0) at (-10.5, 0.75) {$\beta$};
		\node [style=none] (1) at (-9.5, 1.25) {};
		\node [style=none] (2) at (-12.25, 1.25) {};
		\node [style=none] (3) at (-12.25, 0) {};
		\node [style=none] (4) at (-9.5, 0) {};
		\node [style=none] (5) at (-9.5, 1) {};
		\node [style=none] (6) at (-12.25, 1) {};
		\node [style=none, rotate=90] (7) at (-12.5, 0.5) {...};
		\node [style=none, rotate=90] (8) at (-9.5, 0.5) {...};
		\node [style=none] (9) at (-13, 2.5) {};
		\node [style=Z phase dot] (10) at (-12, 2.25) {$\alpha$};
		\node [style=none] (11) at (-10, 2.75) {};
		\node [style=none] (12) at (-13, 2.75) {};
		\node [style=none] (13) at (-10, 1.5) {};
		\node [style=none, rotate=90] (14) at (-10, 2) {...};
		\node [style=none] (15) at (-10, 2.5) {};
		\node [style=none] (16) at (-13, 1.5) {};
		\node [style=none, rotate=90] (17) at (-12.75, 2) {...};
		\node [style=none] (18) at (-9.5, 1.5) {};
		\node [style=none] (19) at (-9.5, 2.75) {};
		\node [style=none] (20) at (-9.5, 2.5) {};
		\node [style=none] (21) at (-13, 1) {};
		\node [style=none] (22) at (-13, 0) {};
		\node [style=none] (23) at (-13, 1.25) {};
		\node [style=none] (24) at (-8.5, 1.5) {$=$};
		\node [style=none, rotate=90] (25) at (-11.25, 1.5) {...};
		\node [style=none] (26) at (-7.5, 2.75) {};
		\node [style=none] (27) at (-4, 2.25) {};
		\node [style=none, rotate=90] (28) at (-7, 1.25) {...};
		\node [style=none] (29) at (-4, 0) {};
		\node [style=none, rotate=90] (30) at (-4.25, 1.25) {...};
		\node [style=none] (31) at (-7.5, 2.25) {};
		\node [style=Z phase dot] (32) at (-5.75, 1.5) {$\ \alpha\!+\!\beta\ $};
		\node [style=none] (33) at (-4, 2.75) {};
		\node [style=none] (34) at (-7.5, 0) {};
		\node [style=none] (35) at (-8.5, 2.25) {\footnotesize $(\bm f)$};
		\node [style=Z phase dot] (36) at (-6.25, -3) {$\ (\textrm{-}1)^a \alpha\ $};
		\node [style=none] (37) at (-4, -1.25) {};
		\node [style=none] (38) at (-8.5, -3) {$=$};
		\node [style=none] (39) at (-7.5, -3) {};
		\node [style=none] (40) at (-4, -4.25) {};
		\node [style=X phase dot] (41) at (-4.75, -2.25) {$a\pi$};
		\node [style=X phase dot] (42) at (-4.75, -4.25) {$a\pi$};
		\node [style=X phase dot] (43) at (-12, -3) {$a \pi$};
		\node [style=none] (44) at (-9.5, -1.5) {};
		\node [style=Z phase dot] (45) at (-11, -3) {$\alpha$};
		\node [style=none, rotate=90] (46) at (-9.75, -3.25) {...};
		\node [style=none] (47) at (-4, -2.25) {};
		\node [style=none, rotate=90] (48) at (-4.75, -3.25) {...};
		\node [style=none] (49) at (-9.5, -2.5) {};
		\node [style=none] (50) at (-12.75, -3) {};
		\node [style=none] (51) at (-9.5, -4.25) {};
		\node [style=X phase dot] (52) at (-4.75, -1.25) {$a\pi$};
		\node [style=none] (53) at (-8.5, -2.25) {\footnotesize $(\bm \pi)$};
		\node [style=X phase dot] (54) at (3, -1.25) {$a\pi$};
		\node [style=none, rotate=90] (55) at (3.75, -3) {...};
		\node [style=X phase dot] (56) at (3, -2.25) {$a\pi$};
		\node [style=Z phase dot] (57) at (-0.75, -2.75) {$\alpha$};
		\node [style=none] (58) at (0.5, -3.75) {};
		\node [style=none] (59) at (1.5, -2.75) {$=$};
		\node [style=none] (60) at (4, -3.75) {};
		\node [style=none] (61) at (0.5, -2.25) {};
		\node [style=none] (62) at (4, -1.25) {};
		\node [style=none, rotate=90] (63) at (0.25, -3) {...};
		\node [style=none] (64) at (4, -2.25) {};
		\node [style=X phase dot] (65) at (3, -3.75) {$a\pi$};
		\node [style=none] (66) at (1.5, -2) {\footnotesize $(\bm c)$};
		\node [style=none] (67) at (0.5, -1.25) {};
		\node [style=X phase dot] (68) at (-1.75, -2.75) {$a\pi$};
		\node [style=hadamard] (69) at (4.25, 2.5) {};
		\node [style=Z phase dot] (70) at (-1, 1.25) {$\alpha$};
		\node [style=none] (71) at (2.25, 1.25) {};
		\node [style=none, rotate=90] (72) at (4.25, 1) {...};
		\node [style=none] (73) at (1.5, 1.25) {$=$};
		\node [style=hadamard] (74) at (4.25, 0.25) {};
		\node [style=none] (75) at (-2.5, 1.25) {};
		\node [style=X phase dot] (76) at (3.25, 1.25) {$\alpha$};
		\node [style=none] (77) at (4.75, 2.5) {};
		\node [style=hadamard] (78) at (-1.75, 1.25) {};
		\node [style=none] (79) at (4.75, 1.75) {};
		\node [style=hadamard] (80) at (4.25, 1.75) {};
		\node [style=none] (81) at (0.5, 0.25) {};
		\node [style=none] (82) at (4.75, 0.25) {};
		\node [style=none, rotate=90] (83) at (0.25, 1) {...};
		\node [style=none] (84) at (0.5, 2.5) {};
		\node [style=none] (85) at (0.5, 1.75) {};
		\node [style=none] (86) at (1.5, 2) {\footnotesize $(\bm{h})$};
		\node [style=Z dot] (87) at (7, 2.5) {};
		\node [style=none] (88) at (6.25, 2.5) {};
		\node [style=none] (89) at (7.75, 2.5) {};
		\node [style=none] (90) at (9.75, 2.5) {};
		\node [style=none] (91) at (10.75, 2.5) {};
		\node [style=none] (92) at (8.75, 3.25) {\footnotesize $(\bm{i1})$};
		\node [style=none] (93) at (8.75, 2.5) {$=$};
		\node [style=none] (94) at (8.75, 0.25) {$=$};
		\node [style=none] (95) at (8.75, 1) {\footnotesize $(\bm{i2})$};
		\node [style=none] (96) at (10.75, 0.25) {};
		\node [style=none] (97) at (7.75, 0.25) {};
		\node [style=none] (98) at (6.25, 0.25) {};
		\node [style=none] (99) at (9.75, 0.25) {};
		\node [style=hadamard] (100) at (6.75, 0.25) {};
		\node [style=hadamard] (101) at (7.25, 0.25) {};
		\node [style=none] (102) at (8.75, -1.75) {\footnotesize $(\bm b)$};
		\node [style=X dot] (103) at (11.75, -3.25) {};
		\node [style=none] (104) at (12.5, -1.75) {};
		\node [style=X dot] (105) at (6, -2.5) {};
		\node [style=Z dot] (106) at (7.25, -2.5) {};
		\node [style=none] (107) at (9.5, -3.25) {};
		\node [style=none] (108) at (5.25, -1.75) {};
		\node [style=none] (109) at (9.5, -1.75) {};
		\node [style=Z dot] (110) at (10.25, -1.75) {};
		\node [style=none] (111) at (8, -1.75) {};
		\node [style=none] (112) at (8.75, -2.5) {$=$};
		\node [style=none] (113) at (5.25, -3.25) {};
		\node [style=X dot] (114) at (11.75, -1.75) {};
		\node [style=none] (115) at (12.5, -3.25) {};
		\node [style=Z dot] (116) at (10.25, -3.25) {};
		\node [style=none] (117) at (8, -3.25) {};
	\end{pgfonlayer}
	\begin{pgfonlayer}{edgelayer}
		\draw [in=-141, out=0, looseness=0.75] (3.center) to (0);
		\draw [in=180, out=-39, looseness=0.75] (0) to (4.center);
		\draw [in=180, out=22, looseness=0.75] (0) to (5.center);
		\draw [in=180, out=39, looseness=0.75] (0) to (1.center);
		\draw [in=0, out=180] (0) to (6.center);
		\draw [in=165, out=0] (2.center) to (0);
		\draw [in=-141, out=0, looseness=0.75] (16.center) to (10);
		\draw [in=180, out=-15] (10) to (13.center);
		\draw [in=180, out=22] (10) to (15.center);
		\draw [in=180, out=39, looseness=0.75] (10) to (11.center);
		\draw [in=0, out=158, looseness=0.75] (10) to (9.center);
		\draw [in=141, out=0, looseness=0.75] (12.center) to (10);
		\draw [bend left] (10) to (0);
		\draw (11.center) to (19.center);
		\draw (15.center) to (20.center);
		\draw (13.center) to (18.center);
		\draw (23.center) to (2.center);
		\draw (21.center) to (6.center);
		\draw (22.center) to (3.center);
		\draw [bend left] (0) to (10);
		\draw [in=-120, out=0] (34.center) to (32);
		\draw [in=180, out=-60] (32) to (29.center);
		\draw [in=180, out=45, looseness=0.75] (32) to (27.center);
		\draw [in=180, out=60] (32) to (33.center);
		\draw [in=0, out=135, looseness=0.75] (32) to (31.center);
		\draw [in=120, out=0] (26.center) to (32);
		\draw [in=180, out=-60, looseness=0.75] (45) to (51.center);
		\draw [in=180, out=45, looseness=0.75] (45) to (49.center);
		\draw [in=180, out=75, looseness=0.75] (45) to (44.center);
		\draw [in=180, out=0, looseness=0.50] (43) to (45);
		\draw (50.center) to (43);
		\draw [in=-60, out=180, looseness=0.75] (42) to (36);
		\draw [in=0, out=180, looseness=0.75] (36) to (39.center);
		\draw [in=180, out=30] (36) to (41);
		\draw [in=60, out=180, looseness=0.75] (52) to (36);
		\draw (37.center) to (52);
		\draw (47.center) to (41);
		\draw (40.center) to (42);
		\draw [in=180, out=-60, looseness=0.75] (57) to (58.center);
		\draw [in=180, out=45, looseness=0.75] (57) to (61.center);
		\draw [in=180, out=75, looseness=0.75] (57) to (67.center);
		\draw [in=180, out=0, looseness=0.50] (68) to (57);
		\draw (62.center) to (54);
		\draw (64.center) to (56);
		\draw (60.center) to (65);
		\draw [in=180, out=-60, looseness=0.75] (70) to (81.center);
		\draw [in=180, out=45, looseness=0.75] (70) to (85.center);
		\draw [in=180, out=75, looseness=0.75] (70) to (84.center);
		\draw [in=180, out=0, looseness=0.75] (78) to (70);
		\draw (75.center) to (78);
		\draw [in=-60, out=180, looseness=0.75] (74) to (76);
		\draw [in=0, out=180, looseness=0.75] (76) to (71.center);
		\draw [in=180, out=45, looseness=0.75] (76) to (80);
		\draw [in=75, out=180, looseness=0.75] (69) to (76);
		\draw (77.center) to (69);
		\draw (79.center) to (80);
		\draw (82.center) to (74);
		\draw (88.center) to (89.center);
		\draw (90.center) to (91.center);
		\draw (98.center) to (97.center);
		\draw (99.center) to (96.center);
		\draw (116) to (114);
		\draw (103) to (110);
		\draw (110) to (114);
		\draw (116) to (103);
		\draw (115.center) to (103);
		\draw (107.center) to (116);
		\draw (110) to (109.center);
		\draw (114) to (104.center);
		\draw [bend right] (113.center) to (105);
		\draw [bend right] (105) to (108.center);
		\draw (105) to (106);
		\draw [bend right] (106) to (117.center);
		\draw [bend left] (106) to (111.center);
	\end{pgfonlayer}
\end{tikzpicture}

%% file: figures/cnot.tikz
\begin{tikzpicture}
	\begin{pgfonlayer}{nodelayer}
		\node [style=Z dot] (0) at (0, 0.5) {};
		\node [style=none] (1) at (1.25, 0.5) {};
		\node [style=X dot] (2) at (0, -0.5) {};
		\node [style=none] (3) at (-1.25, 0.5) {};
		\node [style=none] (4) at (1.25, -0.5) {};
		\node [style=none] (5) at (-1.25, -0.5) {};
	\end{pgfonlayer}
	\begin{pgfonlayer}{edgelayer}
		\draw (0) to (2);
		\draw (0) to (3.center);
		\draw (0) to (1.center);
		\draw (5.center) to (2);
		\draw (2) to (4.center);
	\end{pgfonlayer}
\end{tikzpicture}

%% file: figures/Z-a.tikz
\begin{tikzpicture}
	\begin{pgfonlayer}{nodelayer}
		\node [style=Z phase dot] (0) at (0, 0) {$\alpha$};
		\node [style=none] (1) at (-1.25, 0) {};
		\node [style=none] (2) at (1.25, 0) {};
	\end{pgfonlayer}
	\begin{pgfonlayer}{edgelayer}
		\draw (1.center) to (0);
		\draw (0) to (2.center);
	\end{pgfonlayer}
\end{tikzpicture}

%% file: figures/h-alone.tikz
\begin{tikzpicture}
	\begin{pgfonlayer}{nodelayer}
		\node [style=none] (0) at (-0.25, 0) {};
		\node [style=none] (1) at (1.75, 0) {};
		\node [style=hadamard] (2) at (0.75, 0) {};
	\end{pgfonlayer}
	\begin{pgfonlayer}{edgelayer}
		\draw (0.center) to (2);
		\draw (2) to (1.center);
	\end{pgfonlayer}
\end{tikzpicture}

%% file: figures/X-a-expanded.tikz
\begin{tikzpicture}
	\begin{pgfonlayer}{nodelayer}
		\node [style=X phase dot] (0) at (0, 0) {$\alpha$};
		\node [style=none] (1) at (-1, 0) {};
		\node [style=none] (2) at (1, 0) {};
		\node [style=Z phase dot] (3) at (-4.25, 0) {$\alpha$};
		\node [style=none] (4) at (-5.5, 0) {};
		\node [style=none] (5) at (-3, 0) {};
		\node [style=none] (6) at (-2, 0) {$=$};
		\node [style=hadamard] (7) at (-3.5, 0) {};
		\node [style=hadamard] (8) at (-5, 0) {};
	\end{pgfonlayer}
	\begin{pgfonlayer}{edgelayer}
		\draw (1.center) to (0);
		\draw (0) to (2.center);
		\draw (4.center) to (3);
		\draw (3) to (5.center);
	\end{pgfonlayer}
\end{tikzpicture}

%% file: figures/cz-small.tikz
\begin{tikzpicture}
	\begin{pgfonlayer}{nodelayer}
		\node [style=Z dot] (0) at (-2.75, 0.5) {};
		\node [style=none] (1) at (-1.5, 0.5) {};
		\node [style=X dot] (2) at (-2.75, -0.5) {};
		\node [style=none] (3) at (-4, 0.5) {};
		\node [style=none] (4) at (-1.5, -0.5) {};
		\node [style=none] (5) at (-4, -0.5) {};
		\node [style=hadamard] (6) at (-3.5, -0.5) {};
		\node [style=hadamard] (7) at (-2, -0.5) {};
		\node [style=none] (8) at (-0.75, 0) {$=$};
		\node [style=Z dot] (9) at (1, 0.5) {};
		\node [style=none] (10) at (2, 0.5) {};
		\node [style=Z dot] (11) at (1, -0.5) {};
		\node [style=none] (12) at (0, 0.5) {};
		\node [style=none] (13) at (2, -0.5) {};
		\node [style=none] (14) at (0, -0.5) {};
		\node [style=hadamard] (15) at (1, 0) {};
		\node [style=none] (16) at (2.75, 0) {$=$};
		\node [style=Z dot] (17) at (4.5, 0.5) {};
		\node [style=none] (18) at (5.5, 0.5) {};
		\node [style=Z dot] (19) at (4.5, -0.5) {};
		\node [style=none] (20) at (3.5, 0.5) {};
		\node [style=none] (21) at (5.5, -0.5) {};
		\node [style=none] (22) at (3.5, -0.5) {};
	\end{pgfonlayer}
	\begin{pgfonlayer}{edgelayer}
		\draw (0) to (2);
		\draw (0) to (3.center);
		\draw (0) to (1.center);
		\draw (5.center) to (2);
		\draw (2) to (4.center);
		\draw (9) to (11);
		\draw (9) to (12.center);
		\draw (9) to (10.center);
		\draw (14.center) to (11);
		\draw (11) to (13.center);
		\draw [style=hadamard edge] (17) to (19);
		\draw (17) to (20.center);
		\draw (17) to (18.center);
		\draw (22.center) to (19);
		\draw (19) to (21.center);
	\end{pgfonlayer}
\end{tikzpicture}

%% file: figures/id-simp.tikz
\begin{tikzpicture}
	\begin{pgfonlayer}{nodelayer}
		\node [style=Z phase dot] (0) at (-5.75, 0) {$\alpha$};
		\node [style=Z phase dot] (1) at (-2.75, 0) {$\beta$};
		\node [style=Z dot] (2) at (-4.25, 0) {};
		\node [style=none] (3) at (-1.25, 1) {};
		\node [style=none] (4) at (-1.25, -1) {};
		\node [style=none] (5) at (-7, 1) {};
		\node [style=none] (6) at (-7, -1) {};
		\node [style=none, rotate=90] (7) at (-6.75, 0) {\ldots};
		\node [style=none, rotate=90] (8) at (-1.75, 0) {\ldots};
		\node [style=none] (18) at (0, 0) {=};
		\node [style=Z phase dot] (19) at (2.5, 0) {$\alpha$};
		\node [style=Z phase dot] (20) at (5.5, 0) {$\beta$};
		\node [style=none] (22) at (7, 1) {};
		\node [style=none] (23) at (7, -1) {};
		\node [style=none] (24) at (1.25, 1) {};
		\node [style=none] (25) at (1.25, -1) {};
		\node [style=none, rotate=90] (26) at (1.5, 0) {\ldots};
		\node [style=none, rotate=90] (27) at (6.5, 0) {\ldots};
	\end{pgfonlayer}
	\begin{pgfonlayer}{edgelayer}
		\draw (6.center) to (0);
		\draw (0) to (5.center);
		\draw (1) to (3.center);
		\draw (1) to (4.center);
		\draw [style=hadamard edge] (0) to (2);
		\draw [style=hadamard edge] (2) to (1);
		\draw (25.center) to (19);
		\draw (19) to (24.center);
		\draw (20) to (22.center);
		\draw (20) to (23.center);
		\draw (19) to (20);
	\end{pgfonlayer}
\end{tikzpicture}

%% file: main.bbl
\begin{thebibliography}{10}
\providecommand{\bibitemdeclare}[2]{}
\providecommand{\surnamestart}{}
\providecommand{\surnameend}{}
\providecommand{\urlprefix}{Available at }
\providecommand{\url}[1]{\texttt{#1}}
\providecommand{\href}[2]{\texttt{#2}}
\providecommand{\urlalt}[2]{\href{#1}{#2}}
\providecommand{\doi}[1]{doi:\urlalt{http://dx.doi.org/#1}{#1}}
\providecommand{\bibinfo}[2]{#2}

\bibitemdeclare{misc}{anaconda}
\bibitem{anaconda}
\emph{\bibinfo{title}{{Anaconda: Python for Data Science}}}.
\newblock \bibinfo{note}{\url{https://www.anaconda.com/}}.

\bibitemdeclare{misc}{matplotlib}
\bibitem{matplotlib}
\emph{\bibinfo{title}{{Matplotlib}}}.
\newblock \bibinfo{note}{\url{https://matplotlib.org/}}.

\bibitemdeclare{misc}{numpy}
\bibitem{numpy}
\emph{\bibinfo{title}{{NumPy}}}.
\newblock \bibinfo{note}{\url{https://www.numpy.org/}}.

\bibitemdeclare{inproceedings}{Backens:2015aa}
\bibitem{Backens:2015aa}
\bibinfo{author}{Miriam \surnamestart Backens\surnameend}
  (\bibinfo{year}{2015}): \emph{\bibinfo{title}{Making the stabilizer
  ZX-calculus complete for scalars}}.
\newblock In \bibinfo{editor}{Chris \surnamestart Heunen\surnameend},
  \bibinfo{editor}{Peter \surnamestart Selinger\surnameend} \&
  \bibinfo{editor}{Jamie \surnamestart Vicary\surnameend}, editors: {\sl
  \bibinfo{booktitle}{Proceedings of the 12th International Workshop on Quantum
  Physics and Logic (QPL 2015)}}, {\sl \bibinfo{series}{Electronic Proceedings
  in Theoretical Computer Science}} \bibinfo{volume}{195}, pp.
  \bibinfo{pages}{17--32}, \doi{10.4204/EPTCS.195.2}.

\bibitemdeclare{article}{backens2018zhcalculus}
\bibitem{backens2018zhcalculus}
\bibinfo{author}{Miriam \surnamestart Backens\surnameend} \&
  \bibinfo{author}{Aleks \surnamestart Kissinger\surnameend}
  (\bibinfo{year}{2018}): \emph{\bibinfo{title}{ZH: A Complete Graphical
  Calculus for Quantum Computations Involving Classical Non-linearity}}.
\newblock {\sl \bibinfo{journal}{arXiv preprint arXiv:1805.02175}},
  \doi{10.4204/EPTCS.287.2}.

\bibitemdeclare{article}{horsman2017surgery}
\bibitem{horsman2017surgery}
\bibinfo{author}{Niel \surnamestart de~Beaudrap\surnameend} \&
  \bibinfo{author}{Dominic \surnamestart Horsman\surnameend}
  (\bibinfo{year}{2017}): \emph{\bibinfo{title}{The ZX calculus is a language
  for surface code lattice surgery}}.

\bibitemdeclare{article}{chen2018classical}
\bibitem{chen2018classical}
\bibinfo{author}{Jianxin \surnamestart Chen\surnameend}, \bibinfo{author}{Fang
  \surnamestart Zhang\surnameend}, \bibinfo{author}{Mingcheng \surnamestart
  Chen\surnameend}, \bibinfo{author}{Cupjin \surnamestart Huang\surnameend},
  \bibinfo{author}{Michael \surnamestart Newman\surnameend} \&
  \bibinfo{author}{Yaoyun \surnamestart Shi\surnameend} (\bibinfo{year}{2018}):
  \emph{\bibinfo{title}{Classical simulation of intermediate-size quantum
  circuits}}.
\newblock {\sl \bibinfo{journal}{arXiv preprint arXiv:1805.01450}}.

\bibitemdeclare{inproceedings}{CD1}
\bibitem{CD1}
\bibinfo{author}{B.~\surnamestart Coecke\surnameend} \&
  \bibinfo{author}{R.~\surnamestart Duncan\surnameend} (\bibinfo{year}{2008}):
  \emph{\bibinfo{title}{Interacting quantum observables}}.
\newblock In: {\sl \bibinfo{booktitle}{Proceedings of the 37th International
  Colloquium on Automata, Languages and Programming (ICALP)}},
  \bibinfo{series}{Lecture Notes in Computer Science},
  \doi{10.1007/978-3-540-70583-3\_25}.

\bibitemdeclare{article}{CD2}
\bibitem{CD2}
\bibinfo{author}{B.~\surnamestart Coecke\surnameend} \&
  \bibinfo{author}{R.~\surnamestart Duncan\surnameend} (\bibinfo{year}{2011}):
  \emph{\bibinfo{title}{Interacting quantum observables: categorical algebra
  and diagrammatics}}.
\newblock {\sl \bibinfo{journal}{New Journal of Physics}} \bibinfo{volume}{13},
  p. \bibinfo{pages}{043016}, \doi{10.1088/1367-2630/13/4/043016}.
\newblock \bibinfo{note}{{arXiv:quant-ph/09064725}}.

\bibitemdeclare{conference}{CK}
\bibitem{CK}
\bibinfo{author}{B.~\surnamestart Coecke\surnameend} \&
  \bibinfo{author}{A.~\surnamestart Kissinger\surnameend}
  (\bibinfo{year}{2010}): \emph{\bibinfo{title}{{The compositional structure of
  multipartite quantum entanglement}}}.
\newblock In: {\sl \bibinfo{booktitle}{Automata, Languages and Programming}},
  \bibinfo{series}{Lecture Notes in Computer Science},
  \bibinfo{publisher}{Springer}, pp. \bibinfo{pages}{297--308},
  \doi{10.1007/978-3-642-14162-1\_25}.
\newblock \bibinfo{note}{Extended version: {a}rXiv:1002.2540}.

\bibitemdeclare{book}{CKbook}
\bibitem{CKbook}
\bibinfo{author}{B.~\surnamestart Coecke\surnameend} \&
  \bibinfo{author}{A.~\surnamestart Kissinger\surnameend}
  (\bibinfo{year}{2014}): \emph{\bibinfo{title}{Picturing Quantum Processes}}.
\newblock \bibinfo{publisher}{Cambridge University Press},
  \doi{10.1007/978-3-319-91376-6\_6}.

\bibitemdeclare{article}{qasmpaper}
\bibitem{qasmpaper}
\bibinfo{author}{Andrew~W \surnamestart Cross\surnameend},
  \bibinfo{author}{Lev~S \surnamestart Bishop\surnameend},
  \bibinfo{author}{John~A \surnamestart Smolin\surnameend} \&
  \bibinfo{author}{Jay~M \surnamestart Gambetta\surnameend}
  (\bibinfo{year}{2017}): \emph{\bibinfo{title}{Open quantum assembly
  language}}.
\newblock {\sl \bibinfo{journal}{arXiv preprint arXiv:1707.03429}}.

\bibitemdeclare{conference}{DP2}
\bibitem{DP2}
\bibinfo{author}{R.~\surnamestart Duncan\surnameend} \&
  \bibinfo{author}{S.~\surnamestart Perdrix\surnameend} (\bibinfo{year}{2010}):
  \emph{\bibinfo{title}{{Rewriting measurement-based quantum computations with
  generalised flow}}}.
\newblock In: {\sl \bibinfo{booktitle}{Proceedings of {ICALP}}},
  \bibinfo{series}{Lecture Notes in Computer Science},
  \bibinfo{publisher}{Springer}, pp. \bibinfo{pages}{285--296},
  \doi{10.1007/978-3-642-14162-1\_24}.

\bibitemdeclare{article}{duncan2019graph}
\bibitem{duncan2019graph}
\bibinfo{author}{Ross \surnamestart Duncan\surnameend}, \bibinfo{author}{Aleks
  \surnamestart Kissinger\surnameend}, \bibinfo{author}{Simon \surnamestart
  Pedrix\surnameend} \& \bibinfo{author}{John \surnamestart van~de
  Wetering\surnameend} (\bibinfo{year}{2019}):
  \emph{\bibinfo{title}{Graph-theoretic Simplification of Quantum Circuits with
  the ZX-calculus}}.
\newblock {\sl \bibinfo{journal}{arXiv preprint arXiv:1902.03178}}.

\bibitemdeclare{inproceedings}{FaganDuncan}
\bibitem{FaganDuncan}
\bibinfo{author}{Andrew \surnamestart Fagan\surnameend} \&
  \bibinfo{author}{Ross \surnamestart Duncan\surnameend}
  (\bibinfo{year}{2019}): \emph{\bibinfo{title}{Optimising Clifford Circuits
  with Quantomatic}}.
\newblock In \bibinfo{editor}{Peter \surnamestart Selinger\surnameend} \&
  \bibinfo{editor}{Giulio \surnamestart Chiribella\surnameend}, editors: {\sl
  \bibinfo{booktitle}{{\rm Proceedings of the 15th International Conference on}
  Quantum Physics and Logic (QPL)}}, {\sl \bibinfo{series}{Electronic
  Proceedings in Theoretical Computer Science}} \bibinfo{volume}{287},
  \bibinfo{publisher}{Open Publishing Association}, pp.
  \bibinfo{pages}{85--105}, \doi{10.4204/EPTCS.287.5}.

\bibitemdeclare{phdthesis}{hadzihasanovic2017algebra}
\bibitem{hadzihasanovic2017algebra}
\bibinfo{author}{Amar \surnamestart Hadzihasanovic\surnameend}
  (\bibinfo{year}{2017}): \emph{\bibinfo{title}{The algebra of entanglement and
  the geometry of composition}}.
\newblock Ph.D. thesis, \bibinfo{school}{University of Oxford}.

\bibitemdeclare{inproceedings}{HarnyAmarCompleteness}
\bibitem{HarnyAmarCompleteness}
\bibinfo{author}{Amar \surnamestart Hadzihasanovic\surnameend},
  \bibinfo{author}{Kang~Feng \surnamestart Ng\surnameend} \&
  \bibinfo{author}{Quanlong \surnamestart Wang\surnameend}
  (\bibinfo{year}{2018}): \emph{\bibinfo{title}{Two Complete Axiomatisations of
  Pure-state Qubit Quantum Computing}}.
\newblock In: {\sl \bibinfo{booktitle}{Proceedings of the 33rd Annual ACM/IEEE
  Symposium on Logic in Computer Science}}, \bibinfo{series}{LICS '18},
  \bibinfo{publisher}{ACM}, \bibinfo{address}{New York, NY, USA}, pp.
  \bibinfo{pages}{502--511}, \doi{10.1145/3209108.3209128}.

\bibitemdeclare{article}{heyfron2018efficient}
\bibitem{heyfron2018efficient}
\bibinfo{author}{Luke~Ellis \surnamestart Heyfron\surnameend} \&
  \bibinfo{author}{Earl \surnamestart Campbell\surnameend}
  (\bibinfo{year}{2018}): \emph{\bibinfo{title}{{An efficient quantum compiler
  that reduces T count}}}.
\newblock {\sl \bibinfo{journal}{Quantum Science and Technology}},
  \doi{10.1088/2058-9565/aad604}.

\bibitemdeclare{inproceedings}{SimonCompleteness}
\bibitem{SimonCompleteness}
\bibinfo{author}{Emmanuel \surnamestart Jeandel\surnameend},
  \bibinfo{author}{Simon \surnamestart Perdrix\surnameend} \&
  \bibinfo{author}{Renaud \surnamestart Vilmart\surnameend}
  (\bibinfo{year}{2018}): \emph{\bibinfo{title}{A Complete Axiomatisation of
  the ZX-Calculus for Clifford+T Quantum Mechanics}}.
\newblock In: {\sl \bibinfo{booktitle}{Proceedings of the 33rd Annual ACM/IEEE
  Symposium on Logic in Computer Science}}, \bibinfo{series}{LICS '18},
  \bibinfo{publisher}{ACM}, \bibinfo{address}{New York, NY, USA}, pp.
  \bibinfo{pages}{559--568}, \doi{10.1145/3209108.3209131}.

\bibitemdeclare{misc}{jupyter}
\bibitem{jupyter}
\bibinfo{author}{Project \surnamestart Jupyter\surnameend}:
  \emph{\bibinfo{title}{{Project Jupyter Homepage}}}.
\newblock \bibinfo{note}{\url{https://jupyter.org/index.html}}.

\bibitemdeclare{misc}{tikzit}
\bibitem{tikzit}
\bibinfo{author}{Aleks \surnamestart Kissinger\surnameend}:
  \emph{\bibinfo{title}{{Tikzit Homepage}}}.
\newblock \bibinfo{note}{\url{https://tikzit.github.io/}}.

\bibitemdeclare{article}{kissinger2019cnot}
\bibitem{kissinger2019cnot}
\bibinfo{author}{Aleks \surnamestart Kissinger\surnameend} \&
  \bibinfo{author}{Arianne Meijer-van \surnamestart de~Griend\surnameend}
  (\bibinfo{year}{2019}): \emph{\bibinfo{title}{CNOT circuit extraction for
  topologically-constrained quantum memories}}.
\newblock {\sl \bibinfo{journal}{arXiv preprint arXiv:1904.00633}}.

\bibitemdeclare{inproceedings}{kissinger2012pattern}
\bibitem{kissinger2012pattern}
\bibinfo{author}{Aleks \surnamestart Kissinger\surnameend},
  \bibinfo{author}{Alex \surnamestart Merry\surnameend} \&
  \bibinfo{author}{Matvey \surnamestart Soloviev\surnameend}
  (\bibinfo{year}{2014}): \emph{\bibinfo{title}{Pattern graph rewrite
  systems}}.
\newblock In \bibinfo{editor}{Benedikt \surnamestart L\"owe\surnameend} \&
  \bibinfo{editor}{Glynn \surnamestart Winskel\surnameend}, editors: {\sl
  \bibinfo{booktitle}{{\rm Proceedings 8th International Workshop on}
  Developments in Computational Models, {\rm Cambridge, United Kingdom, 17 June
  2012}}}, {\sl \bibinfo{series}{Electronic Proceedings in Theoretical Computer
  Science}} \bibinfo{volume}{143}, \bibinfo{publisher}{Open Publishing
  Association}, pp. \bibinfo{pages}{54--66}, \doi{10.4204/EPTCS.143.5}.

\bibitemdeclare{misc}{pyzxdoc}
\bibitem{pyzxdoc}
\bibinfo{author}{Aleks \surnamestart Kissinger\surnameend} \&
  \bibinfo{author}{John \surnamestart van~de Wetering\surnameend}:
  \emph{\bibinfo{title}{{PyZX Documentation}}}.
\newblock \bibinfo{note}{\url{https://pyzx.readthedocs.io}}.

\bibitemdeclare{article}{optimisation-paper}
\bibitem{optimisation-paper}
\bibinfo{author}{Aleks \surnamestart Kissinger\surnameend} \&
  \bibinfo{author}{John \surnamestart van~de Wetering\surnameend}
  (\bibinfo{year}{2019}): \emph{\bibinfo{title}{{Reducing T-count with the
  ZX-calculus}}}.
\newblock {\sl \bibinfo{journal}{arXiv preprint arXiv:1903.10477}}.

\bibitemdeclare{article}{kissinger2017MBQC}
\bibitem{kissinger2017MBQC}
\bibinfo{author}{Aleks \surnamestart Kissinger\surnameend} \&
  \bibinfo{author}{John \surnamestart van~de Wetering\surnameend}
  (\bibinfo{year}{2019}): \emph{\bibinfo{title}{Universal {MBQC} with
  generalised parity-phase interactions and {Pauli} measurements}}.
\newblock {\sl \bibinfo{journal}{Quantum}} \bibinfo{volume}{3}, p.
  \bibinfo{pages}{134}, \doi{10.22331/q-2019-04-26-134}.

\bibitemdeclare{inproceedings}{kissinger2015quantomatic}
\bibitem{kissinger2015quantomatic}
\bibinfo{author}{Aleks \surnamestart Kissinger\surnameend} \&
  \bibinfo{author}{Vladimir \surnamestart Zamdzhiev\surnameend}
  (\bibinfo{year}{2015}): \emph{\bibinfo{title}{Quantomatic: A proof assistant
  for diagrammatic reasoning}}.
\newblock In: {\sl \bibinfo{booktitle}{International Conference on Automated
  Deduction}}, pp. \bibinfo{pages}{326--336},
  \doi{10.1007/978-3-319-21401-6\_22}.

\bibitemdeclare{article}{markov2008simulating}
\bibitem{markov2008simulating}
\bibinfo{author}{Igor~L \surnamestart Markov\surnameend} \&
  \bibinfo{author}{Yaoyun \surnamestart Shi\surnameend} (\bibinfo{year}{2008}):
  \emph{\bibinfo{title}{Simulating quantum computation by contracting tensor
  networks}}.
\newblock {\sl \bibinfo{journal}{SIAM Journal on Computing}}
  \bibinfo{volume}{38}(\bibinfo{number}{3}), pp. \bibinfo{pages}{963--981},
  \doi{10.1137/050644756}.

\bibitemdeclare{misc}{maslovreversible}
\bibitem{maslovreversible}
\bibinfo{author}{Dmitri \surnamestart Maslov\surnameend}:
  \emph{\bibinfo{title}{{Reversible Logic Synthesis Benchmarks Page}}}.
\newblock \bibinfo{note}{\url{http://webhome.cs.uvic.ca/~dmaslov/}}.

\bibitemdeclare{article}{nam2018automated}
\bibitem{nam2018automated}
\bibinfo{author}{Yunseong \surnamestart Nam\surnameend},
  \bibinfo{author}{Neil~J \surnamestart Ross\surnameend}, \bibinfo{author}{Yuan
  \surnamestart Su\surnameend}, \bibinfo{author}{Andrew~M \surnamestart
  Childs\surnameend} \& \bibinfo{author}{Dmitri \surnamestart
  Maslov\surnameend} (\bibinfo{year}{2018}): \emph{\bibinfo{title}{Automated
  optimization of large quantum circuits with continuous parameters}}.
\newblock {\sl \bibinfo{journal}{npj Quantum Information}}
  \bibinfo{volume}{4}(\bibinfo{number}{1}), p.~\bibinfo{pages}{23},
  \doi{10.1038/s41534-018-0072-4}.

\bibitemdeclare{inproceedings}{hadzihasanovic2018diagrammatic}
\bibitem{hadzihasanovic2018diagrammatic}
\bibinfo{author}{Kang~Feng \surnamestart Ng\surnameend}, \bibinfo{author}{Amar
  \surnamestart Hadzihasanovic\surnameend} \& \bibinfo{author}{Giovanni
  \surnamestart de~Felice\surnameend} (\bibinfo{year}{2019}):
  \emph{\bibinfo{title}{A diagrammatic calculus of fermionic quantum
  circuits}}.
\newblock \bibinfo{volume}{15}, \bibinfo{publisher}{Episciences.org},
  \doi{10.4230/LIPIcs.FSCD.2018.17}.

\bibitemdeclare{misc}{quipper}
\bibitem{quipper}
\bibinfo{author}{Peter \surnamestart Selinger\surnameend}:
  \emph{\bibinfo{title}{{Quipper Language Homepage}}}.
\newblock \bibinfo{note}{\url{https://www.mathstat.dal.ca/~selinger/quipper/}}.

\bibitemdeclare{unpublished}{euler-zx}
\bibitem{euler-zx}
\bibinfo{author}{Renaud \surnamestart Vilmart\surnameend}
  (\bibinfo{year}{2018}): \emph{\bibinfo{title}{A Near-Optimal Axiomatisation
  of ZX-Calculus for Pure Qubit Quantum Mechanics}}.
\newblock \urlprefix\url{https://arxiv.org/abs/1812.09114}.

\end{thebibliography}
